\documentstyle[twoside,fleqn,espcrc2,psfig]{article}

\newcommand \beq {\begin{equation}}
\newcommand \eeq {\end{equation}}
\newcommand \beqa {\begin{eqnarray}}
\newcommand \eeqa {\end{eqnarray}}

\newcommand{\AmS}{{\protect\the\textfont2
  A\kern-.1667em\lower.5ex\hbox{M}\kern-.125emS}}

\title {Frustration in Lattice Gauge Theory}
\author{
  Vicente Azcoiti \address {Departamento de F\'{\i}sica Te\'orica, 
  Universidad de Zaragoza, 50009 Zaragoza, Spain.},
  Giuseppe di Carlo \address {Istituto Nazionale di Fisica Nucleare, Laboratori
  Nazionali di Frascati, P.O.B. 13, 00044 Frascati, Italy.},
  Eduardo Follana$^a$\thanks{Talk presented by E. Follana.}
	}

\begin{document}

\begin{abstract}

We introduce a U(1) lattice gauge theory which incorporates explicit
frustration in $d>2$. We show, by identifying an
appropiate order parameter and through computer simulations, the
existence of a frustrated region in the phase diagram of the model. We
study this phase diagram and the nature of the transition lines.

\end{abstract}

\maketitle

\section{Introduction}

The importance of frustration and disorder is well known to condensed-matter
physicists working in the field of spin-glasses and related systems 
\cite{Parisi}. In these systems one can find a variety of ``unusual'' phases 
which, in some cases, present completely new phenomena (for example, 
the existence of a multiplicity of vacua not related by a symmetry of the 
action). It is interesting to ask whether similar phenomena could happen
in a field theory and what will be the physical consequences.

As a first step in this direction, we address here the issue of frustration
in lattice gauge theories. Specifically we will study a model in which
frustration is introduced by hand (although inspiration is taken from
the hopping-parameter expansion of QED).

\section{Definition of the model}

The model we have studied presents explicit frustration.
It is defined by the following action:
\beq
S=S_4+S_6=-\beta\sum_{pl}\mathrm{Re}U_{pl}+\beta_6\sum_{p_6}\mathrm{Re}U_{p_6}
\label{action}
\eeq
with $\beta , \beta_6 \geq 0$.
The first part is the standard $U(1)$ pure gauge Wilson action, whereas the 
second part is defined as a sum of contributions over all closed loops 
made up with six (non-repeated) links. These loops can be classified in three 
different classes: planar loops, loops that involve two planes and loops that 
involve three planes. In the special case $d=2$ only the planar loops appear, 
which implies the absence of frustration for this system, consistently with 
the fact that it can be mapped onto the $X-Y$ model within an external field.

\section{Classical ground states and order parameter}

It is not difficult to realize the existence of frustration in this model
at the classical level and in $d\geq 3$, even if we consider only the $S_6$ 
piece. In fact, it is not possible to minimize simultaneously the 
contributions to action (1) of all three classes of loops:
the best we can do is to minimize the planar and three-planes loop 
contributions using some special
chess-board-like configurations, that we will call ``antiferromagnetic'', 
with plaquettes taking values alternatively $\pm 1$. 
The two-planes loop contribution is however not minimized in this way.

We have studied in more detail the case $d=4$. All the results reported here
correspond to this case.
We have checked the relevance of the states described above by doing 
simulations at large values of $\beta_6$. The numerical results show that 
the system tends to freeze in one of these configurations.
This suggests the introduction of a new order parameter for each plane, 
the staggered plaquette, defined as follows:
\beqa
P^s_{\mu\nu} = \frac{1}{V}\sum_x\epsilon(x) \mathrm{Re} U_{\mu\nu} (x)\\
\epsilon(x) = (-1)^{x_1+x_2+x_3+x_4} \nonumber
\eeqa
This order parameter is different from zero in the antiferromagnetic 
vacuum, and vanishes in the ferromagnetic one.

\section{Numerical simulations}

We have performed Montecarlo simulations of this model, using a standard
Metropolis algorithm. 
The observables we measure in the simulation are, in addition to the 
staggered plaquette, the following two quantities:
\beqa
P=\frac{1}{6V}\sum_{pl}\mathrm{Re}U_{pl} \\
P_6=\frac{1}{76V}\sum_{p_6}\mathrm{Re}U_{p_6}
\eeqa
The first one is the usual normalized plaquette, and the second one is
the normalized contribution of the 6-loop part of the action.

\subsection{The line $\beta=0$}

\begin{figure}[!h]
\psrotatefirst
\psfig{figure=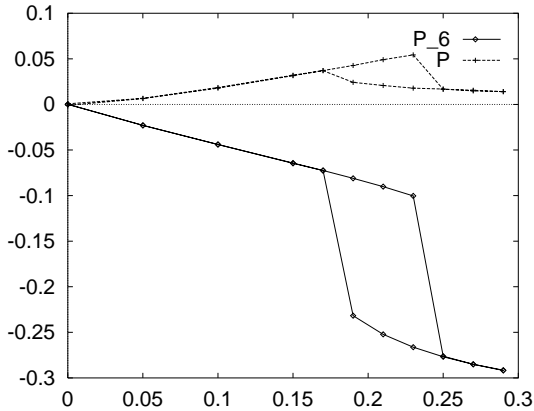,angle=270,width=220pt}
\caption{Plaquette and 6-loop hysteresis cycles at $\beta=0$ against $\beta_6$}
\label{6d4_e}
\end{figure}

\begin{figure}[!h]
\psrotatefirst
\psfig{figure=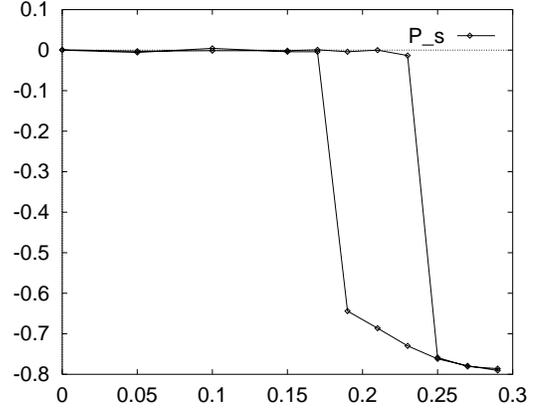,angle=270,width=220pt}
\caption{Staggered Plaquette hysteresis cycle at $\beta=0$ against $\beta_6$}
\label{6d4_ma}
\end{figure}

We show in figures \ref{6d4_e} and \ref{6d4_ma} the annealing cycles
at $\beta=0$. We see a clear hysteresis effect signaling a strong first-order
phase transition. This hysteresis cycle does not show any significant change 
when we increase the simulation time over two orders of magnitude, neither 
when we combine the Metropolis algorithm with an over-relaxation procedure. 

We can see in figure \ref{6d4_ma} that the staggered plaquette is in fact an 
appropiate order parameter for the transition to the antiferromagnetic 
phase.

\subsection{The lines of constant $\beta$}

\begin{figure}[!h]
\psrotatefirst
\psfig{figure=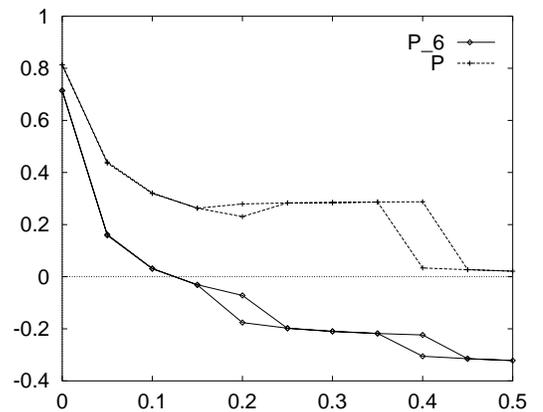,angle=270,width=220pt}
\caption{Plaquette and 6-loop hysteresis cycles at $\beta=1.5$ against 
$\beta_6$}
\label{6d4c4_e}
\end{figure}

\begin{figure}[!h]
\psrotatefirst
\psfig{figure=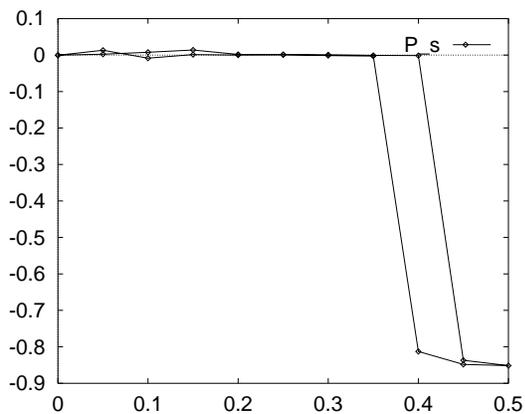,angle=270,width=220pt}
\caption{Staggered Plaquette hysteresis cycle at $\beta=1.5$ against 
$\beta_6$}
\label{6d4c4_ma}
\end{figure}

In figures \ref{6d4c4_e} and \ref{6d4c4_ma} we report the results obtained 
by cycling over $\beta_6$ while keeping a fixed 
$\beta=1.5$, which is well above the value for the usual confined-deconfined
phase transition for the standard $U(1)$ pure gauge theory.
We can clearly see here two transitions: 
the first one corresponds to the continuation of the usual 
confined-deconfined transition (the staggered plaquette remains zero for 
this transition), whereas the second one corresponds to the transition 
to the antiferromagnetic phase, as clearly shown by the jump 
of the staggered plaquette. This results are also stable against 
changes in the Montecarlo time.

\subsection{Phase diagram}

\begin{figure}[!h]
\psrotatefirst
\psfig{figure=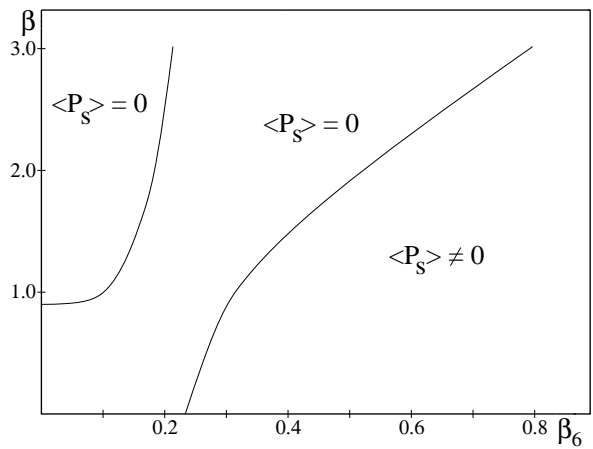,angle=270,width=220pt}
\caption{Phase diagram}
\label{phase_diagram}
\end{figure}

We show in figure \ref{phase_diagram} the tentative phase diagram 
(restricted to the positive quadrant in the $\beta-\beta_6$ plane)
extracted from our simulations of the model described by action (1).
We found three phases separated by two first-order lines
The antiferromagnetic phase is characterized by a non-zero value
of the staggered plaquette, which is zero in the other two phases. In this
phase we have several states, to be precise eight states, related by 
(spontaneously broken) symmetries of the action. The other two phases are the 
continuation of the confined and unconfined phases of the 
standard compact pure gauge model.

\section{Comments}

We have analyzed here the simplest gauge invariant abelian frustrated model and
shown how frustration plays a fundamental role in the dynamics and the vacuum
structure. This work is a first step of the more ambitious program of 
investigating possible implications of frustration in gauge theories with 
dynamical fermions. There are several not very well understood phenomena 
in QED, as the strong coupling phase transition in 3+1 dimensions 
\cite{Kocic,Azcoiti} and the``apparent'' transition in 2+1 dimensions 
\cite{2+1}, the origin of which could be related to the frustrated character 
of the effective fermionic action.

\end{document}